\documentclass[letterpaper,10pt,journal,twocolumn,twoside]{IEEEtran}
\usepackage{times}

\usepackage{epsfig,subfigure,bm,mathrsfs}
\usepackage{tikz}
\usetikzlibrary{decorations.text}
\usepackage{pgfplots}
\usepackage{siunitx}
\usepackage{graphicx,cite}
\usepgfplotslibrary{fillbetween}
\usepackage{nicefrac}
\usepackage{mathtools}
\usepackage{verbatim}
\usepackage{booktabs}
\usepackage[nolist]{acronym}
\usepackage[varg,cmbraces,uprightGreek]{newtxmath}   
\mathtoolsset{showonlyrefs=true}

\makeatletter
\newcommand{\thickhline}{%
	\noalign {\ifnum 0=`}\fi \hrule height 1pt
	\futurelet \reserved@a \@xhline
}
\makeatother

\usepackage{float}
\pgfplotsset{every axis/.append style={
		legend cell align=left,
		ylabel near ticks,
		xlabel near ticks,
		label style={
			font=\footnotesize
		},
		tick label style={font=\footnotesize}
		,legend style={font=\footnotesize}
	}}

\newcommand{\dB}{\ensuremath{\text{\ \si{\deci\bel}}}}

\newcommand{\muinf}{\mathsf{I}}
\newcommand{\Iev}[3]{\muinf^{\mathsf{E,#3}}_{V_{#1}\rightarrow C_{#2}}}
\newcommand{\vecIev}[2]{\bm{\muinf}^{\mathsf{E,#2}}_{\bm{V}\rightarrow {C}_{#1}}}
\newcommand{\Iec}[3]{\muinf^{\mathsf{E,#3}}_{C_{#1}\rightarrow V_{#2}}}
\newcommand{\vecIec}[2]{\bm{\muinf}^{\mathsf{E,#2}}_{\bm{C}\rightarrow V_{#1}}}

\newcommand{\Iapp}[2]{\muinf^{\mathsf{APP},#2}_{#1}}
\newcommand{\fv}[2]{\mathsf{f}^{\mathsf{V}}_{{#1},{#2}}}
\newcommand{\fc}[2]{\mathsf{f}^{\mathsf{C}}_{{#1},{#2}}}

\newcommand{\converge}{\mathscr{D}}

\newcommand{\dv}{d_{\mathrm{v}}}
\newcommand{\dc}{d_{\mathrm{c}}}

\newcommand{\proto}{\mathcal{P}}

\newcommand{\thrMap}{\ensuremath{\gamma^{\mathsf{MAP}}}}
\newcommand{\thrTer}{\ensuremath{\gamma^{\mathsf{TE}}}}
\newcommand{\thrBlock}{\ensuremath{\gamma^{\mathsf{BL}}}}
\newcommand{\thrBP}{\ensuremath{\gamma^\decoding}}

\newcommand{\cw}{\bm{x}}
\newcommand{\obs}{\bm{y}}
\newcommand{\avgsnr}{\gamma}
\newcommand{\capacity}{\mathsf{C}}
\newcommand{\fone}{\mathsf{f_{\mathsf a}}}
\newcommand{\ftwo}{\mathsf{f_{\mathsf b}}}
\newcommand{\snrone}{\gamma_{\mathsf a}}
\newcommand{\snrtwo}{\gamma_{\mathsf b}}
\newcommand{\B}{\bm{B}}
\newcommand{\Z}{\bm{0}}
\newcommand{\decoding}{\mathsf{BP}}
\newcommand{\contour}{\mathscr{B}}

\newcommand{\REVB}{}
\newcommand{\REVA}{}

\begin{document}

\title{\LARGE Improving the Decoding Threshold of Tailbiting Spatially Coupled LDPC Codes by Energy Shaping}
\author{Thomas Jerkovits, \IEEEmembership{Student Member, IEEE}, Gianluigi Liva,  \IEEEmembership{Senior Member, IEEE}, Alexandre Graell i Amat \IEEEmembership{Senior Member, IEEE}
\thanks{T. Jerkovits and G. Liva are with the Institute of Communication and Navigation of the Deutsches Zentrum f\"{u}r Luft- und Raumfahrt (DLR), 82234 Wessling,
   Germany (e-mail: {\tt \{thomas.jerkovits,gianluigi.liva\}@dlr.de}). 
   }
  \thanks{A. Graell i Amat is with the
   Department of Electrical Engineering, Chalmers University of Technology,
   	41296 Gothenburg, Sweden (e-mail: {\tt alexandre.graell@chalmers.se}).}}

\date{\today}

\thispagestyle{empty} \setcounter{page}{1} \maketitle

\begin{abstract}
	We show how the iterative decoding threshold of tailbiting spatially coupled (SC) low-density parity-check (LDPC) code ensembles can be improved over the binary input additive white Gaussian noise channel by allowing the use of different transmission energies for the codeword bits. We refer to the proposed approach as \emph{energy shaping}. We focus on the special case where the transmission energy of a bit is selected among two values, and where a contiguous portion of the codeword is transmitted with  the largest one. Given these constraints,  an optimal  \emph{energy boosting} policy is derived by means of protograph extrinsic information transfer analysis. We show that the threshold of tailbiting SC-LDPC code ensembles can be made close to that of terminated code ensembles while avoiding the rate loss (due to termination). The analysis is complemented by Monte Carlo simulations, which confirm the viability of the approach.		
\end{abstract}



{\pagestyle{plain} \pagenumbering{gobble}}

\begin{acronym}
	\acro{bi-AWGN}{binary input additive white Gaussian noise}
	\acro{LDPC}{low-density parity-check}
	\acro{TB}{tailbiting}
	\acro{TE}{terminated}
	\acro{SC-LDPC}{spatially coupled low-density parity-check}
	\acro{BEC}{binary erasure channel}
	\acro{VN}{variable node}
	\acro{CN}{check node}
	\acro{BP}{belief propagation}
	\acro{MAP}{maximum a posteriori}
	\acro{BICM}{bit interleaved coded modulation}
	\acro{P-EXIT}{protograph extrinsic information transfer}
	\acro{EXIT}{extrinsic information transfer}
	\acro{SNR}{signal-to-noise ratio}
	\acro{LLR}{log-likelihood ratio}
	\acro{APP}{a posteriori probability}
	\acro{MI}{mutual information}
	\acro{UE}{uniform energy}
	\acro{BER}{bit error rate}
	\acro{CER}{codeword error rate}
\end{acronym}
\section{Introduction}

\IEEEPARstart{I}{t} is  known that {for \ac{TE} \ac{SC-LDPC} codes ~\cite{Gallager63,Lentmaier10:CLDPC,KRU13,MLC15}} the \ac{BP} threshold of the underlying code ensemble approaches the \ac{MAP} threshold {in the limit of large coupling lengths}.
{When the coupling length is moderate or small,} the  termination (which is necessary to trigger the decoding wave) entails a non negligible rate loss.
{The rate loss can be avoided by resorting to \ac{TB} \ac{SC-LDPC} codes \cite{HAA+14,HAB+15,CAS+16} at the cost of a (potentially large) threshold degradation. In fact, the \ac{BP} threshold of a   \ac{TB} \ac{SC-LDPC} code ensemble is the same as the one of the underlying uncoupled code ensemble.
	Approaches to improve the \ac{BP} decoding threshold of \ac{TB} \ac{SC-LDPC} code ensembles were introduced in \cite{HAA+14,HAB+15,CAS+16}. They rely  on the possibility of either mapping a specific portion of the codeword to the most reliable bit levels in a bit interleaved coded modulation scheme, or on fixing some of the codeword bits to known values. The latter case, which can be adopted on any binary input channel, still entails a rate loss due to the code shortening. Both approaches aim at triggering the wave-like decoding phenomenon that is at the base of the threshold saturation effect. For the case of  \ac{TE} \ac{SC-LDPC} code ensembles, this is enabled by the stronger protection provided by the termination.} In \cite{Sanatkar16}, a way to mitigate the rate loss is introduced, which relies on appending  \acp{VN} with suitable degree distributions to the \ac{SC-LDPC} graph.

In this letter, we investigate an alternative approach {that} enables large improvements of the \ac{BP} threshold of \ac{TB} \ac{SC-LDPC} code ensembles over the \ac{bi-AWGN} channel. The technique, which preserves the rate of the  \ac{TB} \ac{SC-LDPC} code ensemble, relies on distributing different energies to the codeword bits.  We refer to this approach as \emph{energy shaping}.\footnote{The term ``shaping'' shall not be intended in the sense of signal shaping \cite{Forney84}. As it will be described in Section \ref{preliminaries}, here a time-sharing approach is considered, where the transmission energy is allowed to vary over time.} By doing so, different bit reliabilities are achieved. The ensemble \ac{BP} threshold is analyzed by means of \ac{P-EXIT} \cite{LC07} analysis. Thanks to its capability of dealing with \acp{VN} associated to channels with different \ac{SNR} \cite{Pulini13}, we show how, under the restriction of admitting only two energy values, the boosting level and the length of the boosted portion can be optimized to attain the lowest possible threshold for a given \ac{TB} \ac{SC-LDPC} code ensemble.

{The idea of improving the \ac{BP} threshold of protograph-based \ac{LDPC} code ensembles \cite{Thorpe03}  by means of different transmission energies was originally proposed in \cite{Richter2007}. The approach in \cite{Richter2007} performs a \ac{VN}-by-\ac{VN} optimization over the protograph, where the optimization of the energy allocated to each protograph \ac{VN} is performed through a downhill optimization approach. While very general, the approach may become costly for large protographs. In this letter, we focus  on the simplified case where the transmission energy of a bit can be chosen among two values, and where the largest one is used for the transmission of a contiguous portion of the codeword. The intuition is that, by localizing the \emph{energy boost} on a contiguous portion of the codeword bits, the wave-like decoding effect is triggered. With respect to \cite{Richter2007}, which performs the energy optimization under the assumption that the decoder adopts a parallel message passing schedule, our approach is specifically tailored to operate with the sliding window decoding schedule typically employed in \ac{SC-LDPC} code decoders to reduce the decoding complexity \cite{Lentmaier10:CLDPC}.}

\section{Preliminaries}\label{preliminaries}

We consider transmission over the \ac{bi-AWGN} channel of a (modulated) codeword $\cw=(x_1,\ldots,x_n)$, where $n$ is the block length and  $x_i \in \{ -1, +1 \}$. The channel output is denoted by $\obs=(y_1,\ldots,y_n)$, where 
\[
y_i=\sqrt{f_i} x_i+z_i
\]
with $z_i$ being realizations of independent and identically distributed Gaussian random variables with zero mean and variance $\sigma^2$. Here, $f_i>0$ is a parameter proportional to the energy used for the transmission of $x_i$, such that
\begin{equation}
\frac{1}{n}\sum_{i=1}^n f_i=1. \label{eq:bi}
\end{equation}
We denote the code rate by $R$. The average \ac{SNR}, defined as the ratio between the average energy per information bit $E_b$ and the single-sided noise power spectral density $N_0$, is given by $\avgsnr:={1}/\left(2R\sigma^2\right)$.
We focus on the case where two values for $f_i$ are allowed. We denote them by $\fone$ and $\ftwo$, with $\fone>\ftwo$. The ratio $\phi:=\fone/\ftwo$ is referred to as \emph{boosting factor}. 
We assume next that for the first $\ell$ channel uses the parameter $f_i$ is set to $\fone$, whereas for the remaining $n-\ell$ channel uses $f_i=\ftwo$. We refer to the parameter $\ell$ as the \emph{boosting length}, and by $\lambda:=\ell/n$ as the \emph{normalized boosting length}.
Note that for the first $\ell$ channel uses the \ac{SNR} is $\snrone=\avgsnr \fone$ while for the remaining $n-\ell$ channel uses the \ac{SNR} is $\snrtwo=\avgsnr \ftwo$. Also, observe that \eqref{eq:bi} can now be restated as
$\lambda \fone+(1-\lambda)\ftwo=1$. Hence the transmission parameters are fully specified by the energy shaping parameters $(\phi,\lambda)$. Furthermore, we have that
\begin{equation}
\avgsnr=\lambda\snrone+(1-\lambda)\snrtwo \label{eq:gammaAvg}
\end{equation}
with $\snrone/\snrtwo=\phi$, which yields
\begin{align}
\snrone=\frac{\phi \avgsnr}{\lambda\phi+(1-\lambda)} \qquad \snrtwo=\frac{\avgsnr}{\lambda\phi+(1-\lambda)}.  \label{eq:gamma12}
\end{align}
\REVA{Note that the parameters $(\phi,\lambda)$ together with the average \ac{SNR} $\avgsnr$ are required to determine the \acp{SNR} $\snrone,\snrtwo$.}
Consider a \ac{bi-AWGN} channel with \ac{SNR} $\gamma$, and denote by $\capacity(\gamma)$ its capacity. For a given pair $(\phi,\lambda)$, the capacity is
\begin{equation}
\capacity_{\phi,\lambda}(\avgsnr)=\lambda \capacity(\snrone)+(1-\lambda)\capacity(\snrtwo)\label{eq:cap}
\end{equation}
where the dependency on $\phi$ and $\gamma$ is implicit due to  \eqref{eq:gamma12}.

\subsection{Protograph-Based Spatially Coupled LDPC Codes}

Here, we consider protograph-based \ac{SC-LDPC} codes \cite{MLC15}. In particular, in the following sections we will consider for simplicity a special class of rate-$1/2$ $(\dv,\dc)$ regular \ac{TB} \ac{SC-LDPC} ensembles, where $\dv$ is the \ac{VN} degree and $\dc=2\dv$ is the \ac{CN} degree. A protograph $\proto$ \cite{Thorpe03} is a small bipartite graph comprising a set of $N$ \acp{VN} (also referred to as \ac{VN} types) $\left\{V_1, V_2, \ldots, V_N\right\}$ and a set of $M$ \acp{CN} (i.e., \ac{CN} types)  $\left\{C_1, C_2, \ldots, C_M\right\}$. A \ac{VN} type $V_j$ is connected to a \ac{CN} type $C_k$ by $b_{k,j}$ edges. A protograph can be equivalently represented in matrix form by an $M\times N$ matrix $\B$. The $j$-th column of $\B$ is associated to  \ac{VN} type $V_j$ and the $k$-th row of $\B$ is associated to \ac{CN} type $C_k$. The $(k,j)$ element of $\B$,  $b_{k,j}$, indicates the number of edges connecting $V_j$ and $C_k$. 
A larger graph (derived graph) can be obtained from a protograph by applying a copy-and-permute procedure. The protograph is copied $Q$ times ($Q$ is commonly referred to as lifting factor), and the edges of the different  copies are permuted preserving the original protograph connectivity: If a type-$j$ \ac{VN} is connected to a type-$k$ \ac{CN} with $b_{k,j}$ edges in the protograph, in the derived graph each type-$j$ \ac{VN} is connected to $b_{k,j}$ distinct type-$j$ \acp{CN} (observe that multiple connections between a \ac{VN} and a \ac{CN} are not allowed in the derived graph). The derived graph is the Tanner graph of an \ac{LDPC} code with length $n=NQ$.
A protograph $\proto$ can be used to define a code ensemble $\mathcal{C}^n_{\proto}$. For a given protograph $\proto$, consider all its possible derived graphs with $n$ \acp{VN}. The ensemble $\mathcal{C}^n_{\proto}$ is the collection of codes associated to the derived graphs in the set.

In the following, we consider protograph-based \ac{TB} \ac{SC-LDPC} codes with base matrix in the form
\begin{equation}
\B= \begin{pmatrix*}[l]
\B_0 & \Z   & \Z & \cdots & \B_0 & \B_0 & \cdots & \B_0 \\
\B_0 & \B_0 & \Z & \cdots & \Z   & \B_0 & \cdots & \B_0 \\
\vdots & \vdots & \vdots & \ddots & \vdots  & \vdots & \vdots & \vdots \\
\B_0 & \B_0 & \B_0 & \cdots & \Z  & \Z & \cdots & \Z \\
\Z & \B_0 & \B_0 & \cdots & \Z  & \Z & \cdots & \Z \\
\vdots & \vdots & \vdots & \ddots & \vdots  & \vdots & \vdots & \vdots \\
\Z & \Z & \Z & \cdots & \B_0  & \B_0 & \cdots & \Z\\
\Z & \Z & \Z & \cdots & \B_0  & \B_0 & \cdots & \B_0 \\
\end{pmatrix*}
\label{eq:protograph-base-matrix-tailbiting}
\end{equation}
with $\B_0=(1\,\, 1)$ {and where the number of sub-matrices $\B_0$ per row/column is $\dv$.} Here, $N$ equals $2M$ and it is usually referred to as the number of spatial positions.

\section{Extrinsic Information Transfer Analysis}\label{idea}

\subsection{Analysis over Parallel Bi-AWGN Channels}

The performance of protograph-based \ac{LDPC} codes over parallel channels can be analyzed by the \ac{P-EXIT} analysis. Following \cite{Pulini13}, we consider next the case where the codeword bits corresponding to the $N$ protograph \acp{VN} are transmitted
over $N$  parallel \ac{bi-AWGN} channels. We denote by $\Iev{j}{k}{i}$ the \ac{MI} between the message sent at 
iteration $i$ by the $j$-th \ac{VN} to the $k$-th \ac{CN} and the corresponding codeword bit. Similarly, $\Iec{k}{j}{i}$ denotes the 
\ac{MI} between the message sent at iteration $i$ by the $k$-th \ac{CN} to the $j$-th \ac{VN} and the corresponding codeword bit. We 
further define the \ac{SNR} vector $\bm{\gamma}=(\gamma_1,\gamma_2,\ldots,\gamma_N)$ with $\gamma_j=\snrone$ if $j \in [1,\lambda N]$ and $\gamma_j=\snrtwo$ otherwise. The evolution of the \ac{MI} can be tracked by applying the 
recursion
\begin{align}
\Iev{k}{j}{i}=\fv{k}{j}\left(\vecIec{j}{i-1},\bm{\gamma}\right), \quad
\Iec{k}{j}{i}=\fc{k}{j}\left(\vecIev{k}{i}\right)\label{eq:proto:Ie_general}
\end{align}
with
\[
\vecIec{j}{i}:=\left(\Iec{1}{j}{i},\Iec{2}{j}{i},\ldots,\Iec{M}{j}{i} \right)
\]
and
\[
\vecIev{k}{i}:=\left(\Iev{1}{k}{i},\Iev{2}{k}{i},\ldots,\Iev{N}{k}{i} \right)
\]
where by convention we set $\Iev{j}{k}{i}=\Iec{k}{j}{i}=0$ if $b_{k,j}=0$. In 
\eqref{eq:proto:Ie_general} $\fv{k}{j}$ and 
$\fc{k}{j}$ are the variable and check \ac{EXIT} functions, whose expression can be found in \cite[Sec. IV.A]{Pulini13}.
We finally introduce  $\Iapp{j}{i}$ as the \ac{MI} between the logarithmic \ac{APP} ratio at the $j$-th \ac{VN} in the 
$i$-th iteration and the corresponding codeword bit. 

\subsection{Achievable Regions for Decoding Convergence}

For a $(\dv,\dc)$ regular \ac{TB} \ac{SC-LDPC} ensemble $\mathcal{C}^n_{\proto}$ and for a pair $(\phi,\lambda)$ we say that the \ac{SNR} pair $(\snrone,\snrtwo)$ is \emph{achievable}  under \ac{BP} decoding  if, for a sufficiently large number of iterations and for large $n$, a code picked at random from the $\mathcal{C}^n_{\proto}$ ensemble exhibits (on average) a vanishing small bit error probability, i.e., if $\Iapp{j}{i}$ converges to $1$ for all $j\in \left[1,N\right]$ as $i\rightarrow \infty$. The region of pairs $(\snrone,\snrtwo)$  for which $\Iapp{j}{i}$ converges to $1$  for all 
$j\in \left[1,N\right]$, as $i\rightarrow \infty$, is referred to as the achievable region $\converge(\dv,\dc,\lambda)$. Formally,
\[
\converge(\dv,\dc,\lambda):=\left\{(\snrone,\snrtwo) \bigg| \, \Iapp{j}{i}\rightarrow 1, \, \forall j, \, i\rightarrow \infty   \right\}.
\] 	
For an arbitrary value of $\snrone$, that we denote by $\snrone^\decoding$, one may define the minimum value for $\snrtwo$, that we denote by $\snrtwo^\decoding$,  such that  $(\snrone^\decoding,\snrtwo^\decoding)$ is achievable under \ac{BP} decoding. Note that each pair $(\snrone^\decoding,\snrtwo^\decoding)$ is unequivocally associated with a specific boosting factor as $\phi=\snrone^\decoding/\snrtwo^\decoding$.
The set of pairs $(\snrone^\decoding,\snrtwo^\decoding)$, denoted by $\contour(\dv,\dc,\lambda)$, determines the boundary of the achievable region $\converge(\dv,\dc,\lambda)$.

\subsection{Decoding Thresholds}

\begin{figure}[t]
	\centering
	\includegraphics[width=0.925\columnwidth]{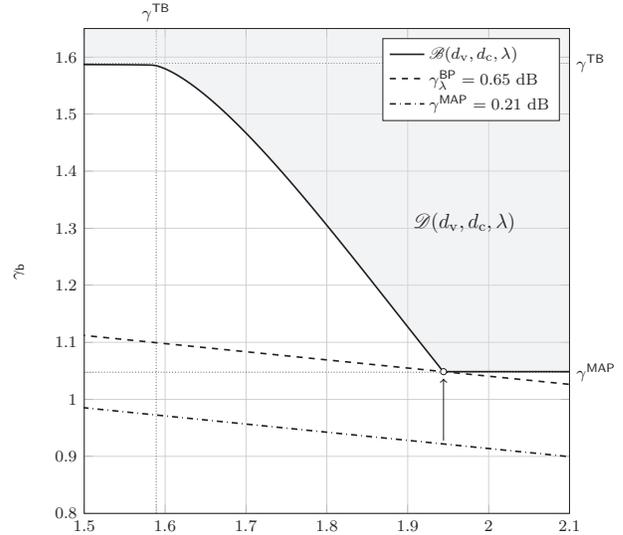}
	\vspace{-7mm}
	\caption{Achievable region for a $(5,10)$ \ac{TB} \ac{SC-LDPC} code ensemble with $N = 128$ and $\lambda = 1/8$.}
	\vspace{-4mm}
	\label{f:awgn_example_N128}
\end{figure}

\begin{figure}[t]
	\centering
	\includegraphics[width=0.90\columnwidth]{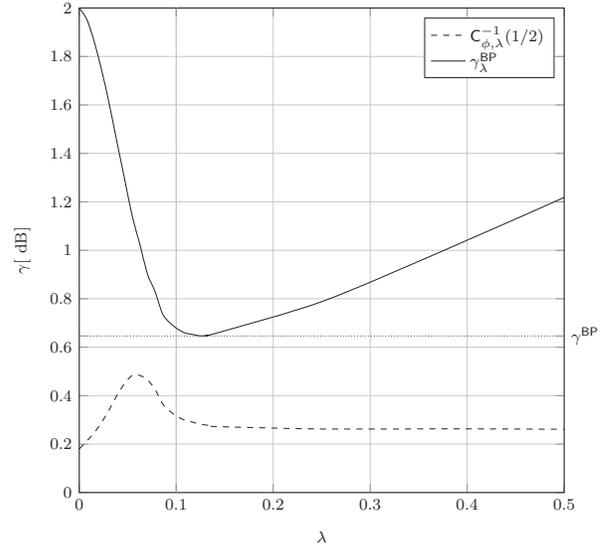}
	\vspace{-3mm}
	\caption{\ac{BP} decoding threshold as a function of $\lambda$, compared with the \ac{SNR} required for rate $R=1/2$, according to \eqref{eq:cap}.}
	\vspace{-4mm}
	\label{f:awgn_threshold_plot_N128}
\end{figure}

The \ac{BP} {decoding threshold} $\gamma_\lambda^\decoding$ of a $(\dv,\dc)$ regular \ac{TB} \ac{SC-LDPC} ensemble when a normalized boosting length $\lambda$  is considered is given by
\begin{equation}
\gamma_\lambda^\decoding := \min_{\substack{(\snrone^\decoding,\snrtwo^\decoding)\in\\ \contour(\dv,\dc,\lambda)}} \Bigg(\lambda\snrone^\decoding+(1-\lambda)\snrtwo^\decoding\Bigg).\label{eq:thr_lambda}
\end{equation} 
In Fig.~\ref{f:awgn_example_N128}, a graphical interpretation of \eqref{eq:thr_lambda} is provided. The figure displays the achievable region $\converge(\dv,\dc,\lambda)$ and its boundary $\contour(\dv,\dc,\lambda)$ for a $(5,10)$  ensemble with $N = 128$ and $\lambda = 1/8$. For a given $\lambda$, one has to find the straight line defined by~\eqref{eq:gammaAvg}, with minimal $\avgsnr$, intersecting the boundary $\contour(\dv,\dc,\lambda)$. For the example in the figure, the minimum is found at $\avgsnr \approx 0.65$~dB, which corresponds to $\snrone \approx 2.89$~dB and $\snrtwo \approx 0.21$~dB, yielding $\phi = 1.85$. By optimizing over the normalized boosting length $\lambda$, we finally find
\begin{equation}
\gamma^\decoding := \min_{\lambda\in [0,1]} \gamma_\lambda^\decoding.
\end{equation}  
Fig.~\ref{f:awgn_threshold_plot_N128} depicts $\gamma_\lambda^\decoding$	as a function of the normalized boosting length for the $(5,10)$ ensemble with $N = 128$. On the same chart, the average \ac{SNR} $\avgsnr$ required to achieve a rate equal to $1/2$ according to \eqref{eq:cap} is depicted. The minimum decoding threshold is attained for $\lambda=1/8$ (with  $\phi = 1.85$). 

\section{Performance Analysis} \label{sec:comparion}

{In the following, results for regular \ac{TB} \ac{SC-LDPC} code (ensembles) with energy shaping are presented through \ac{EXIT} analysis and Monte Carlo simulations. In both cases, windowed decoding \cite{Lentmaier10:CLDPC} has been considered.}

We analyze the performance of the proposed scheme by comparing its iterative decoding thresholds with
\begin{itemize}
	\item[i.] the \ac{MAP} threshold\footnote{\REVA{The estimate of the \ac{MAP} threshold is obtained by exploiting the threshold saturation effect of \ac{SC-LDPC} code ensembles \cite{KRU13}. In particular, we estimate the \ac{MAP} threshold of the uncoupled ensemble by performing an \ac{EXIT} analysis of the \ac{TE} \ac{SC-LDPC} code ensemble for very large $N$ and by setting $\thrMap=\thrTer$.}} $\thrMap$ of the corresponding $(\dv,\dc)$ uncoupled ensemble with \ac{UE};
	\item[ii.] the \ac{BP} threshold $\thrBlock$ of the corresponding $(\dv,\dc)$ uncoupled ensemble with \ac{UE};
	\item[iii.] the \ac{BP} threshold  $\thrTer$ of the corresponding $(\dv,\dc)$ terminated ensemble with \ac{UE}.
\end{itemize}
The analysis is summarized  in Table  \ref{t:awgn_thresholds} for the case of $(\dv,\dc)=(5,10)$. The table provides the thresholds for the different ensembles, for $N=128$, $N=256$, and $N\rightarrow \infty$. For large $N$, the threshold achieved with the proposed approach matches the one of the  $(5,10)$ terminated ensemble with \ac{UE}, saturating to the  \ac{MAP} threshold  of the corresponding $(5,10)$ uncoupled ensemble with \ac{UE}. For moderate-to-small $N$, the proposed approach achieves a larger threshold with respect to the terminated case. However, while for the terminated ensemble the actual coding rate is less than $1/2$ (it reduces to $0.46875$ for $N=128$), the proposed scheme keeps the code rate to the nominal rate. The difference in coding gain between the terminated case and the \ac{TB} \ac{SC-LDPC} code ensemble with energy shaping is limited to about $0.16$~dB for $N=128$, and reduces to $0.09$~dB for $N=256$.
The gain with respect to the uncoupled ensemble threshold exceeds $1.3$~dB for all cases summarized in Table \ref{t:awgn_thresholds}. Table \ref{t:awgn_ens_comp} compares the thresholds achieved by various $(\dv,\dc)$ ensembles for the case of $N=128$. Remarkably, for $N=128$ energy shaping achieves the smallest threshold for the $(4,8)$ ensemble, with a gain of almost $1$~dB over the corresponding block ensemble. This fact has to be attributed to the moderate number  of spatial positions $N$, whereas for $N$ growing large the decoding threshold shall improve with increasing \ac{VN} and \ac{CN} degrees.

\begin{figure}[t]
	\centering
	\includegraphics[width=0.95\columnwidth]{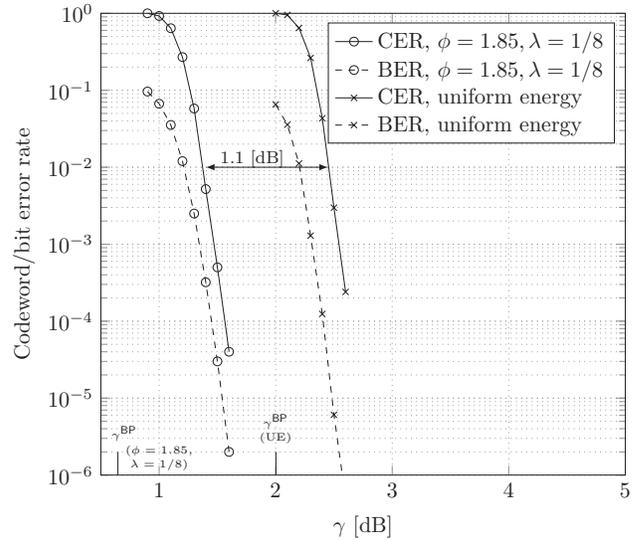}
	\vspace{-4mm}
	\caption{\ac{BER} and \ac{CER} vs. average \ac{SNR} for a $(5,10)$ \ac{TB} \ac{SC-LDPC} code with $N = 128$ and $n=2^{16}$, with and without energy boosting.}
	\vspace{-2mm}
	\label{f:simulation}
\end{figure}

We simulated the performance of the $(5,10)$ \ac{TB} \ac{SC-LDPC} code with $N = 128$, with and without energy boosting. For the case where energy boosting is employed, the boosting parameters have been set to the values minimizing the \ac{BP} threshold, i.e.,  $\lambda=1/8$ (with  $\phi = 1.85$). 
The protograph has been expanded with a lifting factor $Q=512$, yielding a block length $n=2^{16}$.
For decoding, we employed windowed decoding with $50$ iterations per window position (\emph{local} iteration), and a window spanning over $16\times Q$ \acp{VN}. The window is shifted by $2\times Q$ positions at the end of the local iteration steps, circling twice around the tailbiting Tanner graph. The results, in terms of \ac{BER} and \ac{CER}, are provided in Fig.~\ref{f:simulation}. A gain close to $1.1$~dB at $\mathrm{CER}=10^{-2}$ over the \ac{UE} case is achieved.\footnote{\REVB{We may observe that the gain at finite block lengths is smaller than the one predicted by the \ac{EXIT} analysis. In particular, the gain is expected to diminish as the lifting factor $Q$ becomes small.}}

\renewcommand{\arraystretch}{1.3}
\begin{table}[t]
	\centering
	\caption{Comparison of different thresholds for the $(5,10)$ \ac{SC-LDPC} code ensemble and different $N$.}
	\vspace{-2mm}
	\begin{tabular}{c|c|c|c|c} 
		\hline\hline
		$N$      & $\thrBlock$~[dB] & $\thrTer$~[dB]    & $\thrBP$~[dB]          & $\thrMap$~[dB] \\
		\hline
		$128$    & $\num{2.00195}$  & $\num{0.486}$     & $\num{0.646335922977}$ & $\num{0.21}$ \\
		$256$    & $\num{2.00195}$  & $\num{0.343704}$  & $\num{0.431596511332}$ & $\num{0.21}$ \\
		$\infty$ & $\num{2.00195}$  & $\num{0.21}$      & $ \num{0.21}$          & $\num{0.21}$ \\
		\hline\hline
	\end{tabular} 	
	\label{t:awgn_thresholds}
\end{table}

\renewcommand{\arraystretch}{1.3}
\begin{table}[]
	\centering
	\caption{Comparison of different \ac{SC-LDPC} code ensembles and their respective thresholds for $N=128$ and optimal $\lambda$.\textsuperscript{3}}
	\vspace{-2mm}
	\begin{tabular}{c|c|c|c|c}
		\hline\hline
		$(\dv,\dc)$ & $\thrBlock$~[dB]  & $\thrTer (\Delta)$~[dB] & $\thrBP (\Delta)$~[dB] & $\thrMap$~[dB] \\
		\hline
		$(3,6)$ & $\num{1.09718}$ & $\num{0.59285}\,(\num{0.471270155742})$ & $\num{0.599177826673}\,(\num{0.394985711471})$ & $\num{0.454}$ \\
		$(4,8)$ & $\num{1.53824}$ & $\num{0.4606}\, (\num{0.3727847505404})$ & $\num{0.539419360066}\,(\num{0.30098579499})$ & $\num{0.25}$ \\
		$(5,10)$ & $\num{2.00195}$ & $\num{0.486}\,(\num{0.432618934821})$ & $ \num{0.646335922977}\, (\num{0.364855892931})$ & $\num{0.21}$ \\
		\hline\hline
		\multicolumn{5}{l}{\textsuperscript{3}\footnotesize{$\Delta$ is defined as the gap in dB to the corresponding limit according to \eqref{eq:cap}}}
	\end{tabular}
	\vspace{-2mm}
	\label{t:awgn_ens_comp}
\end{table}

\section{Conclusions}\label{conclusions}
We analyzed the convergence behavior of \ac{TB} \ac{SC-LDPC} codes over the \ac{bi-AWGN} channel with different reliabilities assigned to the codeword bits by shaping the transmission energies. By focusing on the case where only two energy levels are allowed, and a contiguous portion of the codeword is transmitted with the largest energy, we showed that large coding gains (e.g., up to $1.1\dB$ for the $(5,10)$ ensemble) can be attained with reference to the case where a uniform energy is employed. The results of the asymptotic analysis are confirmed by finite-length simulations. We conjecture that additional coding gains might be achieved by jointly optimizing the protograph ensemble and the shaping parameters.

\begin{thebibliography}{10}
	\providecommand{\url}[1]{#1}
	\csname url@samestyle\endcsname
	\providecommand{\newblock}{\relax}
	\providecommand{\bibinfo}[2]{#2}
	\providecommand{\BIBentrySTDinterwordspacing}{\spaceskip=0pt\relax}
	\providecommand{\BIBentryALTinterwordstretchfactor}{4}
	\providecommand{\BIBentryALTinterwordspacing}{\spaceskip=\fontdimen2\font plus
		\BIBentryALTinterwordstretchfactor\fontdimen3\font minus
		\fontdimen4\font\relax}
	\providecommand{\BIBforeignlanguage}[2]{{%
			\expandafter\ifx\csname l@#1\endcsname\relax
			\typeout{** WARNING: IEEEtran.bst: No hyphenation pattern has been}%
			\typeout{** loaded for the language `#1'. Using the pattern for}%
			\typeout{** the default language instead.}%
			\else
			\language=\csname l@#1\endcsname
			\fi
			#2}}
	\providecommand{\BIBdecl}{\relax}
	\BIBdecl
	
	\bibitem{Gallager63}
	R.~Gallager, \emph{Low-density parity-check codes}.\hskip 1em plus 0.5em minus
	0.4em\relax Cambridge, MA, USA: MIT Press, 1963.
	
	\bibitem{Lentmaier10:CLDPC}
	M.~Lentmaier, A.~Sridharan, D.~{Costello, Jr.}, and K.~Zigangirov, ``Iterative
	decoding threshold analysis for {LDPC} convolutional codes,'' \emph{{IEEE}
		Trans. Inf. Theory}, vol.~56, no.~10, pp. 5274--5289, Oct. 2010.
	
	\bibitem{KRU13}
	S.~Kudekar, T.~Richardson, and R.~L. Urbanke, ``Spatially coupled ensembles
	universally achieve capacity under belief propagation,'' \emph{{IEEE} Trans.
		Inf. Theory}, vol.~59, no.~12, pp. 7761--7813, Dec. 2013.
	
	\bibitem{MLC15}
	D.~G.~M. Mitchell, M.~Lentmaier, and D.~J. Costello, ``Spatially coupled {LDPC}
	codes constructed from protographs,'' \emph{{IEEE} Trans. Inf. Theory},
	vol.~61, no.~9, pp. 4866--4889, Sep. 2015.
	
	\bibitem{HAA+14}
	C.~H{\"{a}}ger, A.~{Graell i Amat}, A.~Alvarado, F.~Br{\"{a}}nnstr{\"{o}}m, and
	E.~Agrell, ``Optimized bit mappings for spatially coupled {LDPC} codes over
	parallel binary erasure channels,'' in \emph{Proc. {IEEE} Int. Conf. Commun.
		(ICC)}, Sydney, Australia, Jun. 2014, pp. 2064--2069.
	
	\bibitem{HAB+15}
	C.~H\"ager, A.~{Graell i Amat}, F.~Br\"annstr\"om, A.~Alvarado, and E.~Agrell,
	``Terminated and tailbiting spatially coupled codes with optimized bit
	mappings for spectrally efficient fiber-optical systems,'' \emph{J. Lightw.
		Technol.}, vol.~33, no.~7, pp. 1275--1285, Apr. 2015.
	
	\bibitem{CAS+16}
	S.~Cammerer, V.~Aref, L.~Schmalen, and S.~ten Brink, ``Triggering wave-like
	convergence of tail-biting spatially coupled {LDPC} codes,'' in \emph{50th
		Annual Conf. on Information Science and Systems (CISS)}, Princeton (NJ), USA,
	March 2016, pp. 93--98.
	
	\bibitem{Sanatkar16}
	M.~R. Sanatkar and H.~D. Pfister, ``Increasing the rate of spatially-coupled
	codes via optimized irregular termination,'' in \emph{Proc. 9th Int. Symp.
		Turbo Codes and Iterative Inf. Processing}, Brest, France, Sep. 2016, pp.
	121--125.
	
	\bibitem{Forney84}
	G.~Forney, R.~Gallager, G.~Lang, F.~Longstaff, and S.~Qureshi, ``Efficient
	modulation for band-limited channels,'' \emph{{IEEE} J. Sel. Areas Commun.},
	vol.~2, no.~5, pp. 632--647, Sep. 1984.
	
	\bibitem{LC07}
	G.~Liva and M.~Chiani, ``{Protograph LDPC codes design based on EXIT
		analysis},'' in \emph{Proc. {IEEE} Global Telecommun. Conf. (Globecom)},
	Washington (DC), USA, Nov. 2007, pp. 3250--3254.
	
	\bibitem{Pulini13}
	P.~Pulini, G.~Liva, and M.~Chiani, ``{Unequal Diversity {LDPC} Codes for Relay
		Channels},'' \emph{{IEEE} Trans. Wireless Commun.}, vol.~12, no.~11, pp.
	5646--5655, Nov. 2013.
	
	\bibitem{Thorpe03}
	J.~Thorpe, ``Low-density parity-check {(LDPC)} codes constructed from
	protographs,'' NASA JPL, Pasadena, CA, USA, IPN Progress Report 42-154, Aug.
	2003.
	
	\bibitem{Richter2007}
	G.~Richter and M.~Bossert, ``Improving the performance of protograph {LDPC}
	codes by using different transmission energies,'' in \emph{Proc. {IEEE} Int.
		Symp. Inf. Theory (ISIT)}, Nice, France, June 2007, pp. 2251--2255.
	
\end{thebibliography}
\end{document}